\documentclass[preprint2]{aastex}

\begin{document}

\title{HST/NICMOS Observations of the Host Galaxies of Powerful Radio 
Sources: Does Size Matter?}
\author{W.H. de Vries\altaffilmark{1}}
\affil{Inst. of Geophysics \& Planetary Physics, LLNL, Livermore, CA 94550}
\altaffiltext{1}{formerly at Kapteyn Astronomical Institute and
Space Telescope Science Institute.}
\email{devries1@llnl.gov}
\author{C.P. O'Dea}
\affil{Space Telescope Science Institute, 3700 San Martin drive,
Baltimore, MD 21218}
\author{P.D. Barthel}
\affil{Kapteyn Astronomical Institute, P.O. Box 800,
NL-9700 AV, Groningen, The Netherlands}
\author{C. Fanti, R. Fanti}
\affil{University of Bologna, Via Irnerio 46, I-40126, Bologna, Italy}
\and
\author{M.D. Lehnert}
\affil{MPE, Postfach 1603, D-85740, Garching, Germany}


\newcommand{\mmu}{\mbox{$<\!\!\mu\!\!>$}}

\begin{abstract}

We present near-infrared J and K band imaging of a sample of powerful
radio source host galaxies with the Hubble Space Telescope NICMOS2
camera.  These sources have been selected on their double lobed radio
structure, and include a wide range of projected radio source sizes.
The largest projected linear sizes range from the compact Gigahertz
Peaked Spectrum (GPS, $<1$ kpc) and Compact Steep Spectrum (CSS, $<$
20 kpc) radio sources, up to the large-scale ($>$ 20 kpc) classical
doubles (\ion{FR}{2} radio sources). We investigate the dependence of
host galaxy properties (including near-IR surface brightness profiles)
on radio source size, using both our own and published data.  The
absolute magnitudes and surface brightness profiles are consistent
with the host galaxies being regular giant elliptical galaxies rather
than Brightest Cluster Galaxies (BCGs).  We find that the GPS, CSS,
and \ion{FR}{2} host galaxies are a uniform class of objects,
consistent with a scenario in which a powerful radio source evolves
along this size sequence.

\end{abstract}


\section{Introduction}

Powerful radio galaxies play a critical role in our understanding of
both galaxy evolution and the phenomenon of activity in galactic
nuclei. They provide probes of the intracluster medium, particularly
in cooling flow clusters where they reveal the existence of large
scale ordered magnetic fields in the ICM \citep{taylor94}. They
interact with gas in the ISM of their host galaxies producing the
alignment effect and therefore can be used as probes of the content
and structure of the ISM \citep{mccarthy93}. The higher redshift
sources act as beacons which reveal the location of the earliest sites
of the formation of giant ellipticals in the universe (e.g., Spinrad
et al. 1995).  Their luminosities, morphologies, and lifetimes
constrain the processes through which angular momentum is extracted
from gas in the central accretion disk around a supermassive black
hole and jets are formed (e.g., Blandford 1994).  Yet we know little
about the life-cycles of these radio galaxies.

Recent work has identified the GHz Peaked Spectrum (GPS\footnote{While
some GPS sources are known to be larger than the 1 kpc cutoff size, we
only consider here the ones that are smaller than 1 kpc.  Since we
selected our sources to have a double lobed radio structure, our GPS
criterion closely matches the one by Readhead et al.(1996a) for the
Compact Symmetric Objects (CSO's). Throughout this article the GPS and
CSO classifications are considered identical.}) and Compact Steep
Spectrum (CSS) radio sources as likely candidates for the progenitors
of the large-scale powerful classical double (3CR \ion{FR}{2})
sources: e.g., \citet{odea91,odea96,fanti95,snellen96,snellen99,
readhead96a,readhead96b,odea97,odea98,devries98b}. The GPS/CSO and CSS
sources are powerful but compact radio sources whose spectra are
generally simple and convex with peaks near 1 GHz and 100 MHz
respectively. The GPS/CSO sources are entirely contained within the
extent of the narrow line region ($\la$ 1 kpc) while the CSS sources
are contained within the host galaxy ($\la$ 20 kpc).

The typical overall lifetime of a radio source is on the order of
10$^7$ years, significantly shorter than the stellar evolutionary
time-scale (for the bulk of the stars) of a few Gyrs.  The radio
structure therefore most likely originates and expands inside a well
evolved stellar system, and provided we isolate and remove the
instantaneous effects of radio--ISM interaction (like the aligned
component), the global emission properties of these systems should be
identical. Furthermore, within current unification scenarios, radio
galaxies do not permit a direct view into the nucleus at optical and
near-IR wavelengths, as the AGN emission is blocked by obscuring
material. The emission is therefore dominated by the stellar
component, with perhaps minor contributions from dust-reprocessed AGN
light towards the near-IR. Thus, optical and near-IR imaging plays a
powerful role in determining the relationship between the various
classes of radio galaxies.

We have embarked on a multi-wavelength program whose goal is to
understand the relationship between the GPS/CSO and CSS sources on the
one hand and the large-scale sources on the other, and to use the
properties of these objects to constrain radio source evolution. We
have obtained ground-based optical imaging \citep{odea96} which shows
that the GPS/CSO, CSS, and large-scale sources all have similar host
galaxy magnitudes and similar behavior on the $R$ band Hubble diagram
- see also \citet{snellen96}. The host galaxies appear to be passively
evolving ellipticals, with the caveat that the larger sources exhibit
an additional component of blue light associated with the alignment
effect. Our broadband HST images of the CSS sources \citep{devries97}
reveal that they show the alignment effect, even at low redshifts (0.1
$\la z \la$ 0.5). Narrow-band [\ion{O}{3}] imaging of a subset of the
CSS sample shows the CSS radio sources interacting with and exciting
gas as they propagate through the host galaxy and is consistent with
the hypothesis that the CSS are physically young and not confined
sources \citep{devries99,axon00}.

In the near infrared, where host galaxies appear more regular and are
less contaminated by non-stellar components, we obtained ground-based
$J$,$H$, and $K$ band imaging observations of samples of GPS/CSO, CSS,
and large sources matched in redshift and radio flux density (de Vries
et al. 1998a and 1998b, hereafter Paper~I and II respectively).  We
find that (1) the integrated near-IR Spectral Energy Distributions
(SEDs) are similar; (2) the SEDs are consistent with old stellar
populations with formation redshifts between 5 and 10; (3) typically
about 30\% of the K band light is contributed by an additional
component which is well described by emission from dust at a
temperature of $\sim$1000 K. (4) The K band absolute magnitudes are
consistent with passive evolution of the stellar component.

Thus far all our ground-based optical and near-IR observations, as
well as IRAS \citep{heckman94} and ISO \citep{fanti00} observations
suggest that both the host galaxies and AGN are similar in the
GPS/CSO, CSS, and large-scale 3CR \ion{FR}{2}. This continuity of
properties supports the hypothesis that the GPS/CSO sources evolve
through a CSS phase and then into the large-scale 3CR sources.

If this is the case, the number of radio sources per unit size
interval depends on the details of their evolution. Comparing complete
samples of GPS/CSO and CSS sources with the revised 3CR catalog
\citep{laing83}, \citet{odea97} found that the data are consistent
with the proposed evolution scenario if the sources experience strong
luminosity evolution; e.g., the radio luminosity must evolve as
$L_{rad} \propto (size)^{-0.5}$. See also \citet{fanti95, begelman96,
readhead96b, kaiser97, blundell99}.  The required evolution is
consistent with a self-similar linear size evolution of the radio
source in which equipartition is maintained between relativistic
particles and magnetic fields and the radio sources advance at roughly
constant velocity.

In this paper, we utilize the high angular resolution of HST's NICMOS2
camera to determine accurate colors for the stellar and AGN components
separately, and determine well sampled surface brightness profiles.
These unique data allow us to do the following: (1) compare the
stellar colors with the predictions of stellar evolution models; (2)
examine the evolution of the absolute K band magnitudes as a function
of redshift; (3) compare the surface brightness profiles and Kormendy
relations \citep{kormendy85}; and (4) determine the AGN-normalized
radio power as a function of linear size. Our results allow us to
establish the consistency of the host galaxy and AGN properties among
the GPS/CSO, CSS and large scale \ion{FR}{2} radio sources.

\section{Observations and Reduction}

Our sample (cf. Table~1) is drawn from the \citet{odea91} compilation
of GPS sources, and the \citet{heckman94} list of GPS, CSS and
\ion{FR}{2} sources. Each source is selected to have a double-lobe
radio structure, with the possible presence of a radio core. The radio
morphologies among the sample are kept as uniform and symmetric as
possible, in an effort to minimize possible contamination by objects
that appear smaller than they actually are (i.e., foreshortened by
their particular viewing angle).  Furthermore, the sample was matched
among the subclasses in both redshift and radio power. The objects
were observed with the NICMOS 2 camera \citep{thompson98} on board HST
during cycle~7a, as part of the 20 orbit program 7855 (P.I. Chris
O'Dea). In order to sample the extremes of the radio size evolution
better, we chose to image 8 GPS/CSO and 8 \ion{FR}{2}'s, and to use
the remaining 4 orbits on the intermediate size CSS sources. Each
source was observed for 2$\times$640 seconds in the NICMOS filters
F110W and F205W, for a total of 2560 seconds. A few sources with high
declinations had longer orbits (4$\times$704 seconds,
cf. Table~1). The NICMOS F110W and F205W filters are comparable to the
ground-based $J$ and $K$ band filters, though both of the NICMOS
filters are substantially broader -- F110W extends blue-ward to
8000\AA\, while F205W extends up to 2.4 micron.

Each source visit consisted of two dithered pointings, separated by
$\sim$10\arcsec\ along the field-of-view diagonal (27$\farcs$2). Since
all our sources were known to be smaller than 10\arcsec\ (Paper~I)
this dither allowed us to remove remaining sky and detector signatures
to a high degree.

\subsection{Pedestal Removal}

At the time of our reduction, the standard NICMOS reduction pipeline
({\tt calnica} task in STSDAS/IRAF) did not properly remove the
detector background, especially in the filters dominated by the
thermal background (e.g., F205W). This so-called ``pedestal'' of
residual sky\footnote{Actually the pedestal is due to a time varying
DC bias offset, and has nothing to do with actual sky emission. We use
``sky'' in the sense of an unwanted component that has to be removed.}
has to be removed post-pipeline.

The structure of the pedestal in our data was very constant over the
$\sim$6 month period during which our observations in program 7855
were obtained. The fact that we used a large dither and the relative
sparseness of our fields allowed us to construct a ``master sky
dark'', using our own data (with 2$\times$20 input frames for each
filter).  After slight rescaling for each individual source,
subtracting this master sky dark removed almost all of the residuals,
especially in the thermal-dominated F205W filter. For our particular
dataset, this method proved to be more successful in removing the
pedestal than methods by Van der Marel (unpedestal) and Dickinson
(pedsky).

Furthermore, the construction of the mosaic combining both pointings
allowed the removal of ``hot'' and ``cold'' pixels (so-called detector
``grot'') in the overlap region. The final reduced F110W and F205W
images are presented in Figs.~\ref{figure10}--\ref{figure28}.

\subsection{Photometry}

The source photometry was done using the {\tt apphot} package in IRAF.
The magnitudes have been converted to the HST Vega system (NICMOS
handbook 1998), yielding the following zero-points: ZP$_{F110W}=$
22.25, and ZP$_{F205W}$= 21.90. The magnitude is then given by:
m$_{band}$= $-$2.5 log(CountRate) + ZP$_{band}$. The corresponding
Vega fluxes are 1897.3 and 752.2 Jy for F110W and F205W,
respectively. The results are listed in Table~1, columns 6 and 7. The
listed errors are the formal fitting errors (1~$\sigma$), which
include contributions due to Poisson noise, read-out noise, and sky
level uncertainties. In sources with a close companion, we first
subtracted this companion galaxy before doing the photometry
(cf. Sect.~\ref{galsub}). The magnitudes have been corrected for
Galactic extinction, using $A(B)$ values taken from NED\footnote{ The
NASA/IPAC Extragalactic Database (NED) is operated by the Jet
Propulsion Laboratory, California Institute of Technology, under
contract with the National Aeronautics and Space Administration.}
which are based on \ion{H}{1} column densities of Burstein \& Heiles
(1982, 1984). To convert these $A(B)$ values into the proper near-IR
values, we interpolated between the near-IR values given by
\citet{cardelli89}, based on the relative widths and central
wavelengths of F110W and F205W with respect to the $J$ and $K$
bands. This yielded: $A(F110W)=$ 0.380 $\times A(B)$, and $A(F205W)=$
0.104 $\times A(B)$, both under the assumption $R_V=A(V)/E(B-V)=$
3.1. These values differ slightly from the $A(J)$ and $A(K)$ used in
Paper~I.

\subsection{AGN and Galaxy Decomposition}

\begin{figure*}[t]
\epsscale{1.4}
\plotone{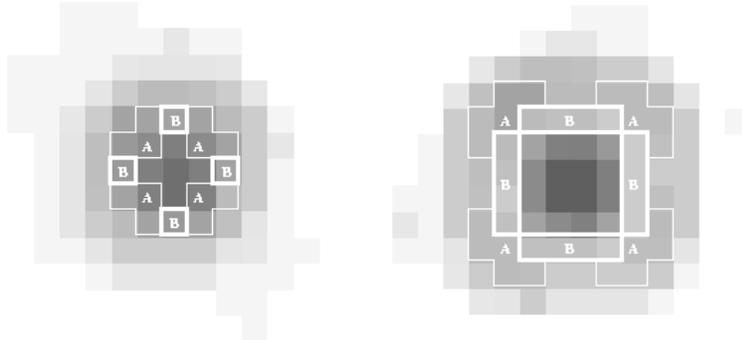}
\caption{Plot of the PSF structure in the F110W (left panel) and
F205W (right panel) filters. To estimate the PS contribution in a
particular source, we use the contrast between the high and low
regions (A and B), as induced by the Airy diffraction pattern.}
\label{psfStruct}
\epsscale{1.0}
\end{figure*}

In order to investigate the properties of the AGN and the stellar
component, we need to decouple them from the total galaxy light.
Fortunately, thanks to the high resolution of the NICMOS camera and
the absence of seeing, we can improve significantly on the
ground-based data (Paper~II) in this respect; though it is still far
from trivial.  Obviously, sources with a strong unresolved component
are decomposed rather easily, however, we find that most sources in
our sample have fairly minor contributions from a nuclear point source
(PS).

An upper bound to the PS contribution can be estimated by assigning
all of the peak nuclear flux to an unresolved component. This peak
nuclear flux is determined by fitting a Gaussian profile to the radial
profile of the source. The Point Spread Function (PSF) is then scaled
up to this value, and the integration of the flux in this scaled PSF
yields the upper limit. This method is applied to the ground based
data (cf. Table~4 of Paper~I), and is used here for comparison
purposes, see Table~3, Columns 2 and 4.

At these near-IR wavelengths, the NICMOS camera slightly resolves the
Airy diffraction pattern. In camera 2, the radii for the first Airy
minimum are approximately 1.8 and 3.1 pixels for F110W and F205W
respectively. This effectively means that up to 16\%\ of the total
flux is spread out over large radii, providing a useful handle on the
total PS flux.  The NICMOS F110W and F205W PSF's are presented in
Fig.~\ref{psfStruct}, where the Airy pattern is readily apparent in
the F205W image. Addition of such a pattern to a galaxy will modulate
the relative fluxes, especially in the areas marked ``A'' and ``B''
where, area ``B'' marks the first Airy minimum. By measuring these
variations in a source and comparing them to those expected in a PSF,
one can calculate the PS contribution to the total galaxy light.  This
method works best for the F205W data, where the Airy pattern is better
resolved, and where the effect of the secondary support structure of
the telescope on the PSF structure is less severe.  PS contributions
recovered this way are listed in Table~3 Columns 3 and 5. Note that
these percentages compare very well to the upper limits derived in the
previous section. Photometry of the separate AGN and galaxy components
is presented in Table~3.

\subsection{Galaxy Isophote Modeling } \label{galsub}

Subtraction of a galaxy model is useful for doing photometry on close
binary galaxies, and in addition may provide insights on deviations
from a simple elliptical galaxy shape revealing structures, such as
line emission due to an aligned component, or very close interacting
galaxies.  We used the {\tt ellipse} package in IRAF to construct
model elliptical galaxies. The major axis position angle and local
ellipticity were allowed to vary. The galaxy center was kept fixed, as
the model otherwise incorporated too much detail: on test-runs,
artificially superposed small galaxies were completely modeled away if
the center was allowed to vary.

Results of the model galaxy subtraction can be found in
Figs.~\ref{figure10}--\ref{figure28}, bottom two panels. It should be
noted that modeling usually failed close to the nucleus, and a point
source is often present in the residual maps. While this seems like a
good way of obtaining the AGN luminosity (cf. last section), negative
features caused by the inaccurate fit compromise accurate photometry.

We also used the galaxy models to determine the galaxy profiles along
the major axis. In essence we determine a galaxy profile by fitting
(at a certain radius) an ellipse that best describes the galaxy (at
that same radius). The local surface brightness, ellipticity, and
position angle are then stored in a table, which is used to extract
the major axis profile and to create the galaxy model at the same
time. A discussion on the fits to these profiles is presented in
\S~\ref{pfits}.

\subsection{Color Maps}

The final step of our reduction process is the creation of
F110W---F205W color maps. We convert counts in the images into Watts
($\nu F_\nu$) before dividing the images. Thus, the ratio represents a
real fractional energy excess. We also smoothed the F110W images to
the resolution of the F205W images using a Gaussian approximation to
the PSF.  However, because of the mismatched Airy patterns in the two
bands, some small artifacts are present in the color images; e.g., the
blue (dark grey) nucleus in 0941$-$080 is surrounded by a red (light
grey / white) ring.  The color maps are plots of the following
function:

\begin{equation} \label{eqn1}
\mbox{ColorMap} = \frac{1 + f \times \mbox{CountRate}_{F110W}
\mbox{[cts/s]}} {1 + \mbox{CountRate}_{F205W}\mbox{[cts/s]}} - 1
\end{equation}

\noindent where CR is the count rate in the image, and the factor
$f=2.68$ is used to normalize the relative energies per count in F110W
and F205W, i.e., one count in F110W represents 2.68 times as much
energy as a count in F205W. The offset by unity in both images
suppresses large color variations in the noise dominated areas of the
map. The color maps are presented in Figs.~\ref{figure10} --
\ref{figure28}, bottom left of the larger panels.

\section{Discussion of Individual Sources}

With reference to Figs.~\ref{figure10}--\ref{figure28}, we note
the following:


{\bf 3C 19} --- \ion{FR}{2} --- This galaxy has an interacting
companion, whose blue color is presumably due to ongoing
star-formation.  The color contrast is rather striking. The radio
structure seems to be oriented towards another (faint) satellite
galaxy.

{\bf 3C 42} --- \ion{FR}{2} --- This source is part of a small group
of galaxies.  Its radio structure is aligned with the closest
companion. No apparent color differences between the various members
are found.

{\bf 3C 46} --- \ion{FR}{2} --- The southern galaxy has a double
nucleus, and most likely contains the radio source. The secondary
nucleus is clearly visible in the residual images. This double nucleus
galaxy is interacting with a slightly smaller companion to the north
(PA 0\degr, distance 1.5\arcsec).

{\bf 3C 67} --- CSS --- In the F110W residual image we see 
a hint of the aligned component which is seen in the WFPC2 F702W image
\citep{devries97}.


{\bf 0404+768} --- GPS/CSO --- This source appears larger in F205W than in
F110W, because of its very red color which cannot be accounted for by
the small estimated Galactic extinction ($A(B)=$ 0.57). Most of the
color, therefore, has to be due to intrinsic absorption.

{\bf 0428+205} --- GPS/CSO --- The galaxy has a red nucleus and is
embedded in a faint halo, extending towards the south and east,
encompassing the small companion to the SSE.

{\bf 0500+019} --- GPS/CSO --- The radio axis points towards the companion
object due south. The accurate centering of the nucleus in the host
galaxy argues against a chance superposition of a galaxy and a
(reddened) quasar, as proposed by \citet{stickel96}.  Thus, we suggest
the red nuclear color is produced in the host galaxy (rather than in
an intervening system -- see also Paper I).


{\bf 3C 200} --- \ion{FR}{2} --- The elongation of the F110W contours
close to the nucleus has the same position angle as the radio jet
\citep{bogers94}.  The associated excess emission is clearly present
in the F110W residual image.

{\bf 0941$-$080} --- GPS --- Like 3C 46, this is an interacting
system. However, there is evidence of enhanced star-formation in the
southern (PA 191\degr, distance 2.6\arcsec) galaxy. The blue
structure is flattened along the major axis of this galaxy, i.e. is
disk-like. The radio source is associated with the large northern
galaxy. Both have blue nuclei.

{\bf 3C 268.3} --- CSS --- Of the prominent line-emission associated
with this source \citep{devries99} only the faintest trace is
recovered in our infrared images. The nucleus is very red, whereas
both companion galaxies have neutral colors.


{\bf 3C 303.1} --- CSS --- The alignment effect can be clearly
recovered in the color map, where, presumably due to H$\alpha$ line
emission in the F110W filter, the aligned structure appears blue.

{\bf 3C 456} --- \ion{FR}{2} --- The F110W residual shows excess
emission towards the south. It is not present in the F205W residual map
though.

{\bf 3C 458} --- \ion{FR}{2} --- This source was not detected, 
possibly due to an error in the published radio source position.  

{\bf 3C 460} --- \ion{FR}{2} --- This is another interacting system
where there are indications of enhanced star formation in the
interacting (PA 195\degr, distance 1.2\arcsec) companion. The host
galaxy exhibits a patch of very red color to the south west of the
nucleus (possibly due to dust).  The radio axis is oriented towards
the general direction of the companion.


{\bf 2352+495} --- GPS/CSO --- This source has boxy isophotes, indicated
by the residuals in the elliptical isophote subtraction.  The radio
axis is perpendicular (in projection) to the major axis of the galaxy.

In summary, the host galaxies of the GPS/CSO, CSS and large
\ion{FR}{2} are morphologically similar. Many show evidence of
interaction with companion galaxies. Some galaxies exhibit red nuclei,
while some companions have blue nuclei. In some galaxies, the
companion galaxy is found along the radio axis, consistent with the
suggestion of \citet{west94} that in prolate galaxies in the centers
of clusters, both the radio source and satellite galaxies will share
the same preferred axis.

\section{Comparison of the GPS/CSO, CSS, and \ion{FR}{2} Hosts}

In this section we compare the host galaxy properties of the different
radio size classes.  As argued in Paper~II, significant differences
between the compact GPS/CSS and the large \ion{FR}{2} host galaxies
might indicate different origins for the radio activity, arguing
against a common evolutionary scenario. In Sect.~\ref{comp2} we
investigate the place of our powerful radio source hosts in the
elliptical galaxy taxonomy.

\subsection{Galaxy and Nuclear Colors}

With the proper decomposition of galaxy and AGN contributions in the
near-IR, we are able to investigate the properties of the colors of
the galaxies as a function of radio source size independent of the AGN
contribution. In addition, the properties of the AGN can be compared
to the radio properties

\begin{figure}[t]
\plotone{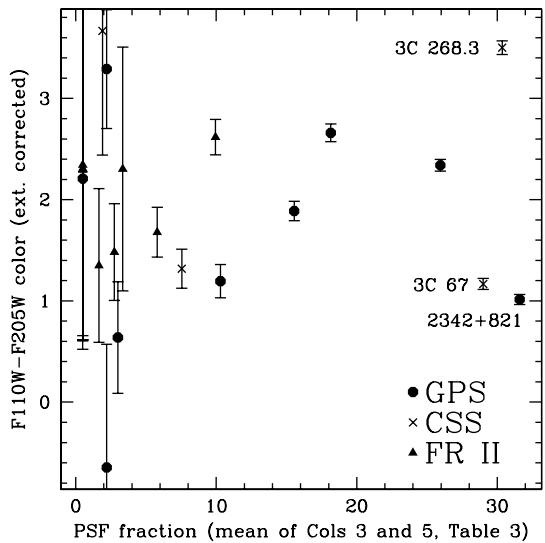}
\caption{Plot of the nuclear color vs. nuclear prominence.  The
nuclear color does not seem to vary with increasing PS dominance. }
\label{NucColor}
\end{figure}

We examine whether there is a correlation between AGN color
(F110W---F205W) and relative AGN prominence in an object. Within the
context of the current unified schemes, e.g., \citet{antonucci93,
urry95}, the AGN should become more dominant in a galaxy as we view it
closer to the axis of the obscuring torus and should become
progressively bluer, as the extinction also decreases. However,
different sources may have differing amounts of obscuration and
extinction unrelated to orientation. Also near-IR emission from a hot
dust component may be present \citep{devries98b}.  The data are
plotted in Fig.~\ref{NucColor}; note that the sources with low PS
dominance have large error bars.  There does not seem to be a
significant trend. Most sources have values of the F110W---F205W color
$\sim 2$, comparable to the host galaxy colors (some exceptions are
discussed below).  This result strengthens the analysis of Paper~II,
where we used galaxy colors which included some AGN contribution.

The absence of a trend in color with nuclear dominance is consistent
with the following. (1) The radio sources are not viewed anywhere near
the opening angle of the obscuring torus and (2) the sources all have
similar amounts of obscuration and/or hot dust emission in our line of
sight to the nucleus.\footnote{The source 2342+821 is an exception,
since it has been re-classified as a quasar \citep{lawrence96}.}

\begin{figure}[t]
\plotone{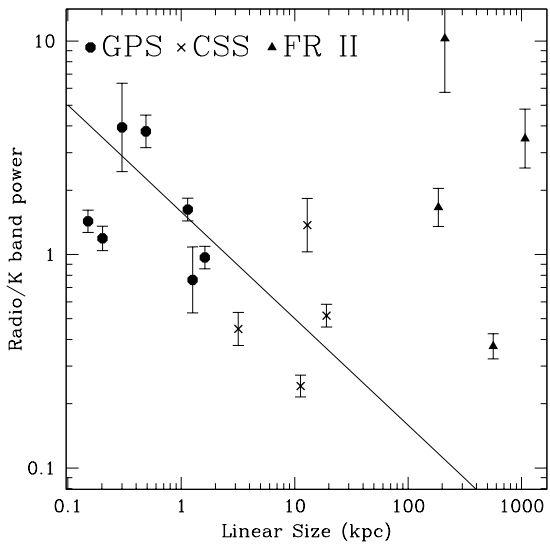}
\caption{Plot of the radio power (at 2.7 GHz in the source rest frame)
normalized by the nuclear K band power (k-corrected) as a function of
radio source projected linear size.  The solid line shows the expected
radio luminosity evolution, with arbitrary scaling. }
\label{SizEvol}
\end{figure}

\subsection{Constraints on Luminosity Evolution of the Radio Sources}

Here we examine whether there is evidence for a decline in radio
luminosity with increasing size. As proposed by \citet{fanti95,
begelman96, odea97}, the luminosity scales as $L_{radio} \propto
(size)^{-0.5}$. If we assume that the radiative output of the AGN is
constant over the radio source lifetimes of ($\sim$10$^{5-7}$ years),
the amount of reprocessed near-IR AGN light should be relatively
constant as well. Therefore, dividing the total radio flux density by
the near-IR flux normalizes the radio power by the intrinsic AGN
radiative luminosity, and makes a direct comparison between the
sources of different sizes possible.  We expect some scatter due to
the following: (1) we measure only the projected linear size rather
than the true radio source size; (2) due to varying obscuration and
redshift, the observed K band nuclear light represents a different
fraction of the true AGN bolometric luminosity; (3) the ratio of AGN
bolometric luminosity to radio source luminosity depends on the
``efficiency'' factor $\epsilon$ which may vary from source to source
or as a function of source size, e.g., \citet{eilek89, chyzy97}; (4)
the radio sources have different radio spectra (though they are all
optically thin at the rest frame frequency of 2.7 GHz used here).

In Fig.~\ref{SizEvol}, we show the k-corrected K-band normalized radio
power versus projected linear size.  The solid line represents the
expected luminosity evolution, arbitrarily scaled.  The GPS/CSO and
CSS sources are consistent with the predicted $L_{radio} \propto
R^{-0.5}$ luminosity evolution. However, the large sources ($R>$ 100
kpc) are inconsistent with this evolution. This plot can be compared
with the P-D diagram in \citet{blundell99} in which they plot
evolutionary tracks for different jet powers, with the main
distinction that our sample has been normalized to the same AGN
output. The slope $\alpha=-0.5$ is consistent with their jet-power
$Q_0=1.3\times10^{45}$ ergs s$^{-1}$ track (at least up to sizes of a
few 100 kpc). Our discrepant large scale sources could fall on a
roughly parallel track for more radio luminous objects, offset to the
GPS/CSO-CSS track. Given the small number of large sources and the
large scatter in the points, this question should be re-addressed with
a larger sample. If this tentative result is confirmed, it would have
significant implications for radio galaxy evolution.  The current plot
does not exclude the possibility that the AGN radiative energy output
(i.e., $K$ band light) drops in the older and larger sources; the
ratio of radio to $K$ band luminosity would therefore increase as the
sources age relative to the predictions of a constant AGN output
model.  Another possibility is that radio luminosity decline slows
down in the larger sources, contrary to the \citet{kaiser97}
prediction. This might be the case if the ambient density profile
flattens \citep{odea97,kaiser97}.

\subsection{Stellar Population Synthesis Models}

Since the galaxy light in the near-IR is due to the stellar population
of the host, we can use stellar synthesis models \citep{bruzual00} to
compare actual and expected colors as a function of mean population
age and metallicity (cf. Paper~II). For this purpose we generated
F110W and F205W magnitudes and colors with $R,J,H,$ and $K$ band
filters, since the HST filters are not included in the standard code.

\begin{figure}[t]
\plotone{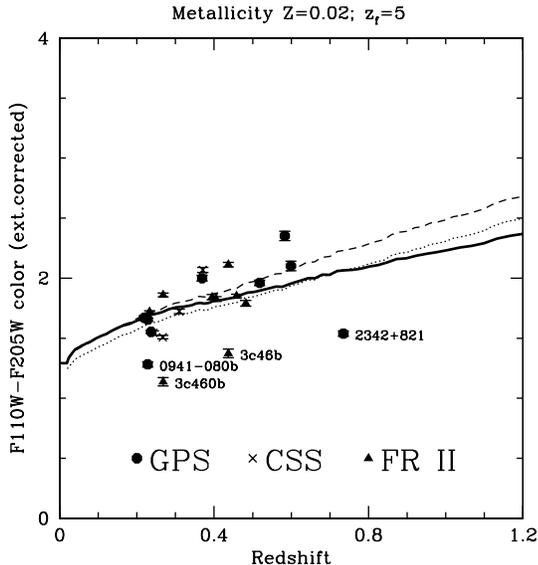}
\caption{Plot of J-K color vs. redshift. The solid line represents
colors expected from a passively evolving system, with a formation
redshift of 5 and a mean metallicity of solar. The short and long
dashed lines are colors of non-evolving galaxies, with mean galaxy
ages of 5 and 10 Gyr, respectively. Note the bluer than expected
colors for all three of the close companions (designated by the letter
b). The GPS/CSO 2342+821 might be too blue because it is a quasar. }
\label{JcolPlot}
\end{figure}

\begin{figure}[t]
\plotone{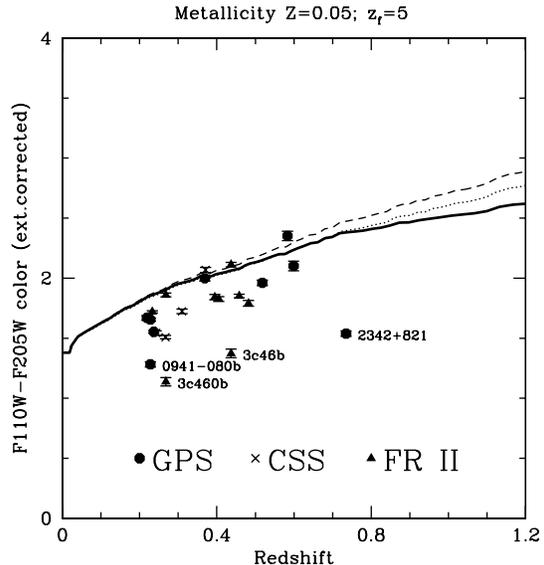}
\caption{Similar to  Fig.~\ref{JcolPlot}, but with a mean metallicity of
2.5$\times$ solar.}
\label{KcolPlot}
\end{figure}

The results are presented in Figs.~\ref{JcolPlot} and
\ref{KcolPlot}. The colors are based on the AGN-contribution
subtracted magnitudes, and should therefore be due to the underlying
stellar population exclusively. We plotted various evolutionary
tracks, similar to those in Figs.~4 and 5 of Paper~II. The solid track
represents the case where an instantaneous starburst at the formation
redshift ($z_f=$ 5) was allowed to passively evolve with time. The
mean metallicity of the model galaxy was kept fixed at Z = 0.02 (left
panel, the solar value), and Z = 0.05 (2.5 times solar, right
panel). The short and long dashed lines are the non-evolving cases,
which implies all sources are observed at the same stage of their
evolution.  The mean intrinsic age of the galaxy is assumed to be 5
Gyr for the bottom dashed track, and 10 Gyr for the top. Consistent
with results from Paper~II, we find the sources in our sample are best
represented by a metallicity slightly above the solar value and an old
mean population age ($\ga$ 5 Gyr). This population age implies a high
formation redshift, and well evolved populations.  Thus, these radio
sources are born and evolve inside mature elliptical galaxies.

The host galaxies of GPS/CSO, CSS, and \ion{FR}{2} sources appear to
have similar near-IR colors, confirming the results of
Paper~II. However, the higher resolution allow us to separate out
close companions, and, as can be seen in Figs.~\ref{JcolPlot} and
\ref{KcolPlot}, all three of these close companions are significantly
bluer than their larger neighbors. This blue excess is most likely due
to ongoing star-formation, triggered by the interaction with their
more massive companion. The color maps show this blue excess as rather
localized regions in 0941$-$080b and 3C 460b; while in 3C 46b it is
more evenly distributed over the galaxy. The presence of active
star-formation implies these small companion galaxies are relatively
gas-rich, unlike their more massive neighbors.  The remaining other
source which appears too blue is GPS/CSO 2342+821. This source is
probably a quasar, also based on its position in Fig.\ref{NucColor}.

In summary, the host galaxies of GPS/CSO, CSS, and \ion{FR}{2} radio
galaxies have similar near-IR colors. They are consistent with stellar
populations with a metallicity slightly above the solar value and an
old mean population age ($\ga$ 5 Gyr).  In the next section we will
compare our sample as a whole to other types of luminous ellipticals.

\section{Comparison of the Host Galaxies with other Ellipticals}\label{comp2}

Elliptical galaxies can be characterized by just a few parameters: the
mean surface brightness \mmu, effective radius $R_e$, and central
velocity dispersion $\sigma$. Ellipticals populate a plane in this 3D
space: the fundamental plane, e.g., \citet{djorgovski87}. Since we
currently lack velocity dispersions for these galaxies, we will use
the projection of the fundamental plane onto the (\mmu,$R_e$)
plane. This correlation has been investigated first by
\citet{kormendy77, kormendy85}, and is now known as the Kormendy
relation. Since the absolute luminosities of our galaxies, due to the
small redshift range, are comparable, the variation in host galaxy
size is expected to be equally small. As a consequence, the variation
in central velocity dispersion among our sample galaxies should also
be small, making the \mmu -- $R_e$ correlation informative
(cf. \S~\ref{pfits}).

\subsection{Evolution of the Absolute Host Galaxy Magnitude} 

A plot of absolute luminosity versus redshift has some nice
features. First, sources are compared on the same absolute (physical)
scale, given the proper k-correction, and not on apparent quantities.
Furthermore, plotting against redshift provides a baseline over which
the data can be compared with evolutionary tracks based on stellar
synthesis models.  And finally, a comparison to published data from
the literature on different types of elliptical galaxies is possible.

\begin{figure}[t]
\plotone{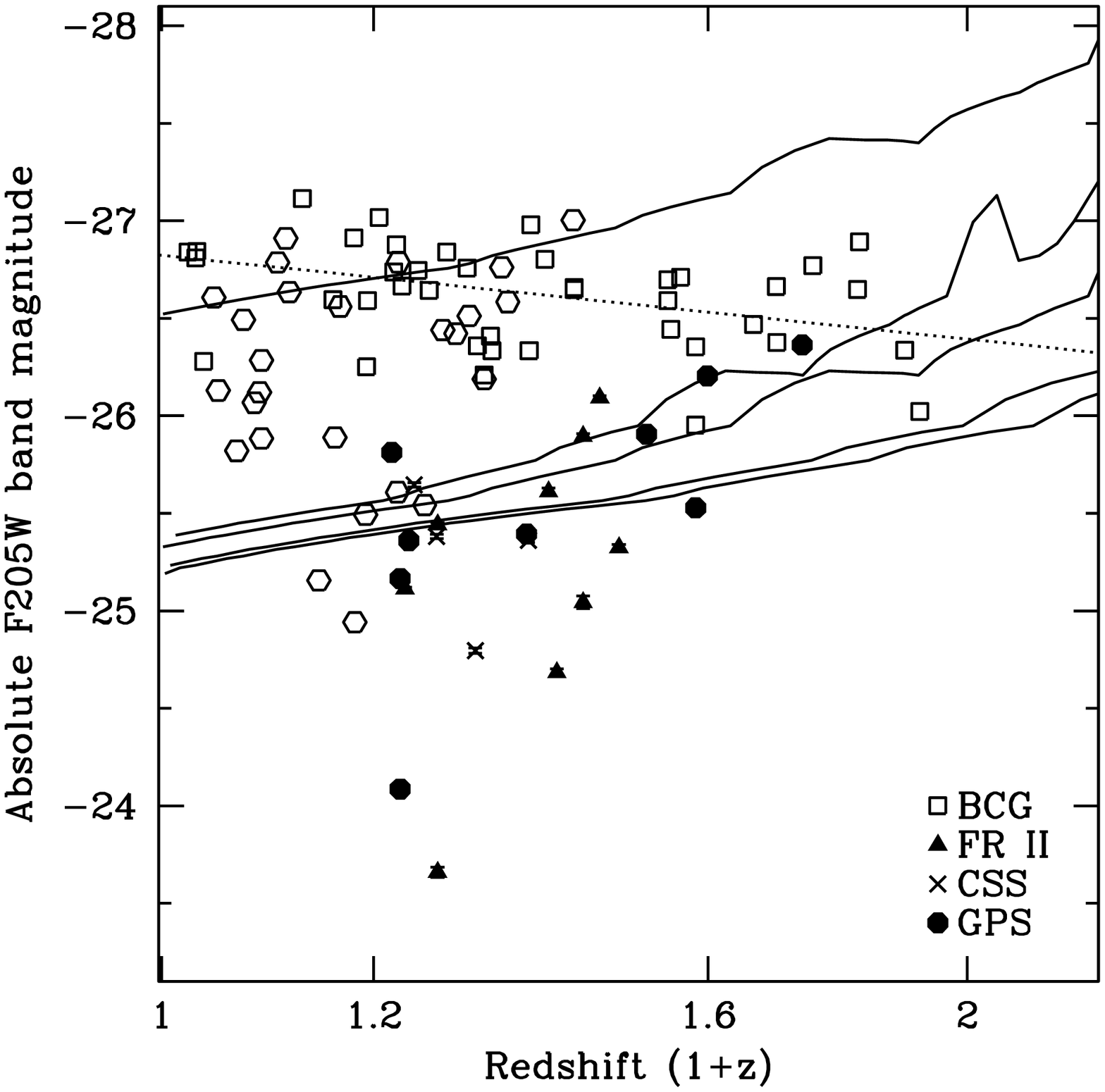}
\caption{Absolute F205W magnitude versus redshift.  BCG data are taken
from \citet{aragon93} and from \citet{collins98}, after being
converted from $K$ into F205W magnitudes. Open squares are BCGs with
an X-ray luminosity in excess of 3$\times$10$^{44}$ ergs s$^{-1}$,
while open hexagons are BCGs with lower X-ray luminosity. The bottom
set of 4 tracks are passive evolutionary track for $5\times10^{11}$
M$_\odot$ luminous mass, with a mean metallicity of $\sim$solar and
formation redshifts ($z_f$) of 1.5, 2, 5, and 8 (top to bottom)
respectively. The top solid track is for $1.5\times10^{12}$ M$_\odot$
luminous mass and $z_f=2$. The dashed track is a best fit to the
bright BCG sample, roughly consistent with a non-evolutionary
scenario.}
\label{absMag}
\end{figure}

We carried out this analysis on our ground-based data in Paper~II.
However, the poorer resolution of the ground-based data resulted in
incomplete AGN--galaxy decomposition, e.g., 1404+286 and 1345+125 are
known to be contaminated by emission from their nuclei.  In
Fig.~\ref{absMag}, we plot the absolute magnitude of the galaxy,
without the AGN contribution. As in Paper~II, data from the literature
on (usually radio quiet) Brightest Cluster Galaxies (BCGs) have been
included, taken from \citet{aragon93} and \citet{collins98}. The
literature $K$-band data have been converted into the F205W filter
system.  Note that these data have not been decomposed into galaxy and
AGN components like our NICMOS data. However, given the fact that BCGs
generally exhibit flat, ``cuspy'', luminosity profiles near their
centers, e.g., \citet{hoessel87, faber97, graham96}, a steep PS-like
component is most probably an insignificant contribution to the
overall luminosity in these galaxies.

As in Fig.~6 of Paper~II, the brightest BCGs (which also tend to be
the more X-ray luminous objects) appear to have a different
evolutionary history than our sample and the predicted tracks for
passively evolving galaxies.  The bright upper end of our sample is
consistent with a passive evolutionary scenario, more so than seen in
Paper~II in the ground based data-set, where in some cases the AGN
contribution was significant. The fact that our sample covers a range
of intrinsic galaxy masses is reflected in the increase in scatter
towards lower redshifts.  The lack of similar galaxies at somewhat
higher redshifts is a selection effect. Something similar seems to
apply to the lower X-ray luminosity sources, i.e. it is uncertain
whether those sources are really absent at higher redshifts.
Fig.~\ref{absMag} has some important implications.  First, our hosts
are less luminous than BCGs at similar redshifts.  \citet{owen89} and
\citet{owen91} find a similar result for Classical Doubles
(\ion{FR}{2}'s), which are significantly fainter than BCGs. Second,
the redshift evolution of our sample seems more consistent with
passive evolution than that of the brightest BCG's, which appear to
increase in luminous mass (possibly through merging) as they
evolve. And finally, we see no dependence of absolute magnitude on
radio source size.

\subsection{Surface Brightness Profiles} \label{pfits}

\begin{figure}[t]
\plotone{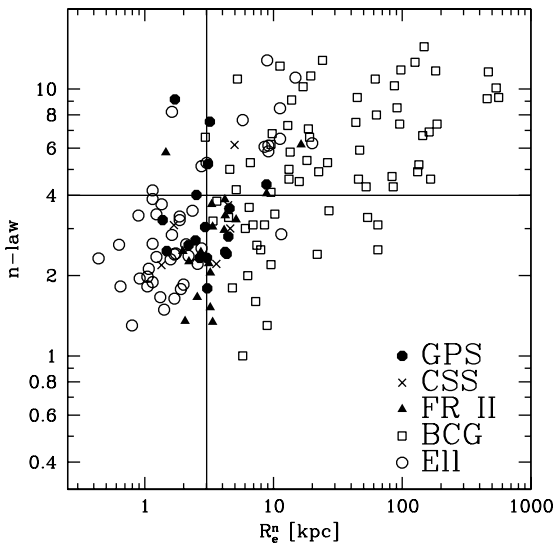}
\caption{Plot of the exponent $n$ of the $R^{1/n}$ law versus the
corresponding effective radius for both the F110W and F205W data.  We
include data from the literature on BCGs \citep{graham96} and a
sample of ellipticals, ranging from small to large \citep{caon93}. The
lines divide the sample into ``ordinary'' ($R_e<3$kpc,$n<4$) and
``bright'' ($R_e>3$kpc,$n>4$) ellipticals (lower left and upper right
quadrant).}
\label{nlaw}
\end{figure}

Luminosity profiles and their derived quantities -- effective radius
($R_e$), effective surface brightness ($\mu_e$), mean surface
brightness (\mmu), and their general shape are all important
quantities for discriminating between various objects. Spiral galaxies
usually have exponential profiles, whereas ellipticals are better
fitted with de Vaucouleurs' type laws ($\mu(r) \propto
R^{1/n}$). While the morphological differences between spirals and
ellipticals are obvious\footnote{when S0's are not taken into account}
and ellipticals as a class appear rather similar, they do exhibit
significant differences in their profiles (cf. Fig.~\ref{nlaw}).  For
instance, \citet{faber97} find a correlation between absolute
luminosity and profile shape, in the sense that the more massive
systems have shallower inner profiles. Less luminous ellipticals have
profiles which lack this ``core'', and remain steep all the way up the
resolution limit.  Unfortunately, the physical size scale of this cusp
has been found to be on the order of 500 pc, e.g., \citet{lauer95,
faber97}.  This ``break radius'' corresponds to about 1 pixel for our
sources (given the redshift range), which is clearly not enough to
detect flattening on these scales. We can, however, fit the profile
with a generalized de Vaucouleurs' law ($R^{1/n}$). This exponent $n$
also correlates with elliptical type. Following \citet{graham96}, we
define:

\begin{equation} \label{eqnP}
\mu(R) = \mu_0 + \frac{2.5 b_n}{\ln(10)} \left(\frac{R}{R_e}\right)^{1/n}
\end{equation}

\noindent with $r_e$ the scale radius, $\mu_0$ the central surface
brightness, and the constant $b_n$ approximated by: $b_n \approx 2n -
0.327$. Note that in the $n=1$ case the profile is exponential, and
better represents spirals.  All our source major axis luminosity
profiles were fitted with this law, and a $\chi^2$ minimization
yielded the best fitting values, given in Table~4.

\citet{caon93} found a correlation between the effective radius and
the exponent $n$, in the sense the galaxies with the smaller $R_e$'s
also have steeper profiles (lower values of $n$). They divided the
elliptical family into ``ordinary'' ($R_e <$ 3 kpc, $n <$ 4), and
``bright'' ($R_e >$ 3 kpc, $n >$ 4) classes. \citet{graham96} added
BCG data to the same plot, confirming the trend. BCGs were found to
belong to the ``bright'' class, and actually have the highest values
of $n$ \citep{graham96}. The data on our sources are overplotted on
the literature data in Fig.~\ref{nlaw}. The GPS/CSO, CSS and large
\ion{FR}{2} are consistent with having similar values of effective
radius and $n$.  Most of our sample lies between the ``ordinary'' and
``bright'' regions of parameter space.  None of the host galaxies in
our sample have an exponential ($n=1$) profile, and therefore the
radio sources do not seem to reside in spiral hosts.  There is an
offset between BCGs and our sample objects, i.e., the BCGs tend to
have larger effective radii and shallower profiles than the radio
source host galaxies.

\subsection{The Kormendy Relation}

\begin{figure}[t]
\plotone{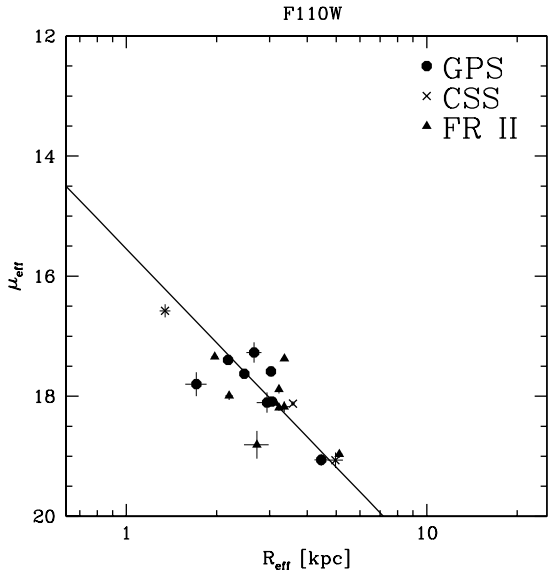}
\caption{Kormendy relation for our sample in the F110W filter.
$R_e$ and $\mu_e$ were determined using the
generalized $R^{1/n}$ law.}
\label{kormRel}
\end{figure}

\begin{figure}[t]
\plotone{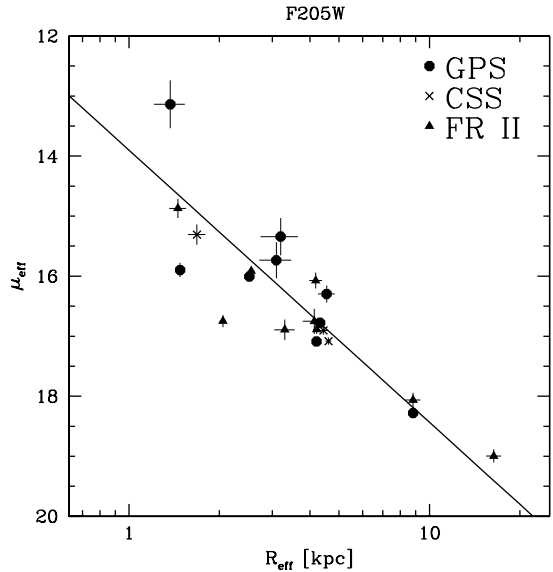}
\caption{Same as Fig.~\ref{kormRel}, but for the F205W filter. Note
the similar slopes and different intercepts.}
\label{kormRel2}
\end{figure}

The slope of the $\mu_e$ -- $R_e$, or Kormendy relation has been found
to vary with the mean physical size of the sample under
consideration. Initially, \citet{kormendy85} established the slope of
the relation as $\alpha \approx 2.1$, where the fit is made to $\mu_e
\propto \alpha\log(R_e)$, and $\mu_e$ is on the magnitude scale. His
sample included a large fraction of (local) small ellipticals and
bulges, and not many giant galaxies. Shortly afterwards,
\citet{hoessel87} noted a significant difference in slope between BCGs
and non-BCGs, in the sense non-BCGs display a steeper slope ($\alpha
\approx$ 4.5, compared to $\alpha \approx$ 3.1 for BCGs). They
suggested, among other things, a different formation process for BCGs
compared to field ellipticals.  Perhaps more relevant to our radio
galaxies, various studies on \ion{FR}{1} and \ion{FR}{2} galaxies find
a similar discrepancy between the two. \citet{smith89} find
\ion{FR}{2}'s to have steeper profiles than BCGs, while the
\ion{FR}{1} slope is consistent with the BCG slope. Another telling
find is that \ion{FR}{2}'s are found to have a lower absolute
luminosity (in $V$) than BCGs (true for F205W as well,
cf. Fig.~\ref{absMag}), while \ion{FR}{1}'s are
comparable. \citet{ledlow95} find a steepening in slope going from a
sample of radio cD galaxies, to BCGs, to non-BCGs ($\alpha =$ 3.3 --
3.8 -- 4.0).

In this sense, it is interesting to determine the slope of the
galaxies in our sample.  Even though our data are in the near-IR and
the data in the literature range from $B$ to $R$ band, the variations
in slope as a function of color are found to be rather small; i.e.,
the zero-points vary, while the slopes do not (e.g., Pahre et
al. 1996).  In Figs.~\ref{kormRel} and \ref{kormRel2}, we plotted the
Kormendy relations for F110W and F205W respectively.  We corrected the
effective surface brightness for the $(1+z)^4$ cosmological
dimming. The fits have slopes of $\alpha =$ 5.2$\pm$0.6 (F110W), and
$\alpha =$ 4.0$\pm$0.5 (F205W) for a de Vaucouleurs' law ($n=$ 4).
For the generalized $R^{1/n}$ law the values are: $\alpha =$
5.2$\pm$0.5 (F110W), and $\alpha =$ 4.5$\pm$0.4 (F205W).

These values strongly suggest our sample to consist of non-BCG
sources. Combined with evidence from the literature on differences
between powerful radio sources (\ion{FR}{2}'s) and BCGs (e.g., Owen \&
Laing 1989), differences between \ion{FR}{1} and \ion{FR}{2} sources
(e.g., Smith \& Heckman 1989), and similarities between \ion{FR}{1}
and BCGs \citep{ledlow95,smith89}, we conclude that GPS/CSO and CSS
hosts are similar to \ion{FR}{2} hosts but significantly different
from at least some \ion{FR}{1} hosts.  This implies GPS/CSS radio
galaxies in general do not evolve into BCG-\ion{FR}{1} type sources.
However, it is clear that not all \ion{FR}{1}s live in BCGs (e.g.,
Ledlow \& Owen 1995).

It will be important to expand this study to include larger numbers of
objects, including \ion{FR}{1} host galaxies.  If future work should
suggest that GPS/CSS sources do not evolve into \ion{FR}{1}s this
would imply the following: either \ion{FR}{1} progenitors are not
GPS/CSO sources when they are young, or, they are {\em faint} GPS/CSO
sources and therefore do not enter the powerful radio samples we are
studying. In the latter case the \ion{FR}{1} sources would have much
less radio luminosity evolution than \ion{FR}{2}'s.

\section{Summary and Conclusions}

We present HST/NICMOS observations of 19 powerful radio galaxies
spanning a wide range of radio source size and including the GPS/CSO,
CSS, and large \ion{FR}{2} radio galaxies. We have obtained F110W (J)
and F205W (K) band imaging in the NICMOS2 camera.  Our results are the
following:

\begin{itemize}

\item{Based on the accurate decomposition of the stellar and AGN
components of the host galaxy, we find these radio galaxy hosts are
not dominated by a nuclear point source in the near-IR.  The nuclear
J-K colors do not depend on radio source size or the prominence of the
nuclear point source.  Thus, any material which obscures the AGN is
optically thick at these near-IR wavelengths in all three classes of
source.  }

\item{The host galaxies of the GPS/CSO, CSS and large \ion{FR}{2} are
morphologically similar. Many show evidence of interaction with
companion objects at differing distances, ranging from the binary
nucleus in the southern component of 3C 46, to much wider pairs like
3C 16 and 0941$-$080. Seven galaxies exhibit red nuclei, while some
companion galaxies have blue nuclei (companions to 3C 19, 3C 46, 3C
460, 0941$-$080). In nine galaxies, a companion object is found along
the radio axis, consistent with the suggestion of \cite{west94} that
in prolate galaxies in the centers of clusters, both the radio source
and satellite galaxies will share the same preferred axis.}

\item{The colors of the host galaxy stellar population are in
agreement among the radio source classes and are consistent with those
of a passively evolving old stellar population.  The mean inferred
population age and metallicity are $\sim$5 Gyr and solar,
respectively.}

\item{The absolute magnitude at K band (F205W) is similar for the
GPS/CSO, CSS, and \ion{FR}{2} host galaxies. The dependence of
absolute magnitude on redshift is consistent with simple passive
evolution models. On the other hand, a comparison sample of bright
BCGs exhibits (1) brighter absolute magnitudes, and (2) perhaps more
complex evolution suggesting the accumulation of light (mass) as the
galaxies evolve.}

\item{Due to the high angular resolution of HST, we were able, for the
first time, to determine meaningful radial profiles of these sources,
despite their typical optical sizes of a few arc seconds. The observed
range of effective radii and surface brightness profile slopes are in
agreement among the three classes of radio source and are consistent
with L$_*$ ellipticals and are inconsistent with those of BCGs. The
behavior of the galaxies on the Kormendy relation is also consistent
with these results.  }

\item{We normalized the radio luminosity by the AGN K band luminosity
in order to determine the radio source luminosity evolution as a
function of projected source size.  The GPS/CSO and CSS sources are
consistent with the predicted $L_{radio} \propto R^{-0.5}$ luminosity
evolution. However, the large sources ($R>$ 100 kpc) are inconsistent
with this evolution. Given the small number of large sources and the
large scatter in the points, this question should be re-addressed with
a larger sample. If this tentative result is confirmed, it would have
significant implications for radio galaxy evolution.  One possibility
is that the AGN radiative energy output (i.e., $K$ band light) drops
in the older and larger sources; then the ratio of radio to $K$ band
luminosity would increase as the sources age relative to the
predictions of a constant AGN output model.  Another possibility is
that radio luminosity evolution slows down in the larger sources. This
might be the case if the ambient density profile flattens
\citep{odea97}. }

\end{itemize}

In summary, the host galaxies of GPS/CSO, CSS, and \ion{FR}{2}
galaxies form a uniform class of objects, consistent with the radio
source evolutionary scenario.  There is some tentative evidence that
the GPS/CSO and CSS sources follow the expected radio luminosity
evolution, while the very large sources do not.

\acknowledgments

We thank Bob Becker, Mark Lacy, and Anton Koekemoer for their careful
reading of the manuscript.  This work was partially supported by NASA
through grant GO-07855.01-96A from the Space Telescope Science
Institute, which is operated by the Association of Universities for
Research in Astronomy, Inc., under NASA contract NAS 5-26555. Part of
WDV's work was performed under the auspices of the U.S. Department of
Energy by University of California Lawrence Livermore National
Laboratory under contract No. W-7405-Eng-48.

\clearpage

\begin{deluxetable}{lllllll}
\tablenum{1}
\tablecaption{Photometry results}
\tablehead{
  \colhead{Source} & \colhead{ID} & \colhead{$z$} & \colhead{A(B)\tablenotemark{a}} &
  \colhead{exp.time} & \colhead{F110W mag\tablenotemark{b}} & 
  \colhead{F205W mag\tablenotemark{b}} 
}
\startdata
3C 16     & FR2 & 0.405 & 0.17 & 2$\times$640 & 18.81$\pm$0.01 & 16.98$\pm$0.02 \\
3C 19     & FR2 & 0.482 & 0.28 & 2$\times$640 & 18.49$\pm$0.02 & 16.71$\pm$0.02 \\
3C 42     & FR2 & 0.395 & 0.22 & 2$\times$640 & 17.82$\pm$0.01 & 15.99$\pm$0.01 \\
3C 46     & FR2 & 0.437 & 0.21 & 2$\times$640 & 18.06$\pm$0.01 & 15.94$\pm$0.01 \\
3C 46b\tablenotemark{c}& FR2 & 0.437 & 0.21 & 2$\times$640 & 18.14$\pm$0.02 & 16.76$\pm$0.03 \\
3C 67     & CSS & 0.310 & 0.30 & 2$\times$640 & 17.55$\pm$0.01 & 15.99$\pm$0.01 \\
3C 93.1   & CSS & 0.243 & 0.77 & 2$\times$640 & 16.36$\pm$0.01 & 14.84$\pm$0.01 \\
0404+768  & GPS/CSO & 0.599 & 0.58 & 2$\times$704 & 18.25$\pm$0.03 & 16.04$\pm$0.01 \\
0428+205  & GPS/CSO & 0.219 & 1.70 & 2$\times$640 & 16.18$\pm$0.02 & 14.51$\pm$0.01 \\
0500+019  & GPS/CSO & 0.583 & 0.29 & 2$\times$640 & 18.97$\pm$0.03 & 16.62$\pm$0.01 \\
0710+439  & GPS/CSO & 0.518 & 0.30 & 2$\times$640 & 18.07$\pm$0.02 & 16.12$\pm$0.01 \\
3C 200    & FR2 & 0.458 & 0.08 & 2$\times$640 & 17.63$\pm$0.01 & 15.79$\pm$0.01 \\
0941$-$080  & GPS & 0.228 & 0.08 & 2$\times$640 & 16.87$\pm$0.01 & 15.24$\pm$0.01 \\
0941$-$080\tablenotemark{c}& GPS & 0.228 & 0.08 & 2$\times$640 & 17.62$\pm$0.01 & 16.34$\pm$0.02 \\
3C 268.3  & CSS & 0.371 & 0.03 & 2$\times$640 & 17.98$\pm$0.01 & 15.49$\pm$0.01 \\
1323+321  & GPS & 0.369 & 0.02 & 2$\times$640 & 17.91$\pm$0.02 & 15.99$\pm$0.01 \\
3C 303.1  & CSS & 0.267 & 0.08 & 2$\times$704 & 16.88$\pm$0.01 & 15.34$\pm$0.01 \\
3C 456    & FR2 & 0.233 & 0.08 & 2$\times$640 & 17.01$\pm$0.01 & 15.20$\pm$0.01 \\
3C 458\tablenotemark{d}& FR2 & 0.289 & 0.19 & 2$\times$640 & 20.73$\pm$0.06 & 19.51$\pm$0.08 \\
3C 460    & FR2 & 0.268 & 0.17 & 2$\times$640 & 17.19$\pm$0.01 & 15.37$\pm$0.01 \\
3C 460b\tablenotemark{c}& FR2 & 0.268 & 0.17 & 2$\times$640 & 18.25$\pm$0.02 & 17.04$\pm$0.03 \\
2342+821  & GPS/CSO & 0.735 & 0.78 & 2$\times$704 & 17.62$\pm$0.01 & 16.24$\pm$0.01 \\
2352+495  & GPS/CSO & 0.237 & 0.71 & 2$\times$640 & 16.66$\pm$0.01 & 15.15$\pm$0.01 \\
\enddata
\tablenotetext{a}{Value based on Burstein \& Heiles HI maps.}
\tablenotetext{b}{One $\sigma$ errors.}
\tablenotetext{c}{Interacting companion galaxy.}
\tablenotetext{d}{Most likely misidentified.}
\end{deluxetable}

\begin{deluxetable}{lrll|lrll|lrll}
\tablenum{2}
\tablecaption{Radio data on sample}
\label{tab2}
\tablehead{
  \colhead{Source} & \colhead{P.A.\tablenotemark{a}} & \colhead{Size\tablenotemark{b}} & \colhead{Ref} & 
  \colhead{Source} & \colhead{P.A.\tablenotemark{a}} & \colhead{Size\tablenotemark{b}} & \colhead{Ref} & 
  \colhead{Source} & \colhead{P.A.\tablenotemark{a}} & \colhead{Size\tablenotemark{b}} & \colhead{Ref}
}
\startdata
3C 16    & $+$35 &  75 & 1  & 0428+205 & $-$24 & 0.25& 6  & 3C 303.1  & $-$50 & 2.5  & 9  \\ 
3C 19    & $+$30 &  12 & 2  & 0500+019 & $-$7  & 0.03& 7  & 3C 456    & $+$17 & 12   & 10 \\
3C 42    & $-$49 &  36 & 3  & 0710+439 &    0  & 0.03& 8  & 3C 460    & $+$36 & 6    & 11 \\
3C 46    & $+$68 & 150 & 4  & 3C 200   & $-$28 & 25  & 1  & 2342+821  & $-$70 & 0.20 & 6  \\
3C 67    & $-$5  &   4 & 5  &0941$-$080& $-$32 & 0.06& 7  & 2352+495  & $-$18 & 0.08 & 12 \\
3C 93.1  & $-$19 & 0.25& 6  &3C 268.3  & $-$20 & 2   & 9  &           &       &      &    \\
0404+768 & $+$45 & 0.15& 6  &1323+321  & $-$47 & 0.08& 6  &           &       &      &    \\
\enddata
\tablenotetext{a}{Position angle, measured North through East.}
\tablenotetext{b}{Radio angular size in arcseconds.}
\tablerefs{1) Bogers et al. 1994, 2) Jenkins et al. 1977, 3)
Leahy et al. 1991, 4) Neff et al. 1995, 5) Sanghera et al. 1995, 6)
Dallacasa et al. 1995, 7) Dallacasa et al. 1998, 8) Fey et al. 1996,
9) Akujor et al. 1995, 10) Harvanek \& Hardcastle 1998,
11) McCarthy et al. 1995 + McCarthy et al. 1991, 12) Polatidis et
al. 1995}
\end{deluxetable}

\begin{deluxetable}{lrrrrllcc}
\tablenum{3}
\tablecaption{Source decompositions and Photometry}
\tablehead{
  \colhead{} & \multicolumn{2}{c}{F110W\tablenotemark{a}} & 
  \multicolumn{2}{c}{F205W\tablenotemark{a}} &
  \colhead{} &\colhead{} &\colhead{} &\colhead{} 
\\
  \colhead{Source} & \colhead{\%$^{\mbox{max}}_{\mbox{psf}}$} & \colhead{\%$^{\mbox{act}}_{\mbox{psf}}$} &
  \colhead{\%$^{\mbox{max}}_{\mbox{psf}}$} & \colhead{\%$^{\mbox{act}}_{\mbox{psf}}$} &
  \colhead{F110W$_{\mbox{gal}}$} & \colhead{F205W$_{\mbox{gal}}$} &
  \colhead{F110W$_{\mbox{nuc}}$} & \colhead{F205W$_{\mbox{nuc}}$}
}
\startdata
3C 16     &  10 &  1 & 18 &  1 & 18.81$\pm$0.01 & 16.98$\pm$0.02 &    23$\pm$2   &    21$\pm$2 \\
3C 19     &   6 &  1 & 11 &  1 & 18.50$\pm$0.02 & 16.71$\pm$0.02 &    23$\pm$2   &    21$\pm$2 \\
3C 42     &   6 &  2 & 10 &  1 & 17.84$\pm$0.02 & 16.00$\pm$0.02 &  22.1$\pm$0.8 &  20.7$\pm$1.2 \\
3C 46     &   4 &  1 &  7 &  1 & 18.06$\pm$0.01 & 15.94$\pm$0.01 &    23$\pm$2   &    21$\pm$2 \\
3C 46b\tablenotemark{b}&  7 &  3 & 19 & 3 & 18.17$\pm$0.01 & 16.80$\pm$0.03 &  22.1$\pm$0.7 &  20.6$\pm$0.7 \\
3C 67     &  27 & 34 & 41 & 24 & 18.01$\pm$0.01 & 16.28$\pm$0.01 & 18.72$\pm$0.08& 17.55$\pm$0.11 \\
3C 93.1   &   7 &  8 & 15 &  7 & 16.46$\pm$0.01 & 14.91$\pm$0.01 & 19.07$\pm$0.29& 17.75$\pm$0.34 \\
0404+768  &  24 & 14 & 37 & 22 & 18.42$\pm$0.04 & 16.31$\pm$0.01 & 20.35$\pm$0.17& 17.69$\pm$0.12 \\
0428+205  &   4 &  2 &  7 &  3 & 16.20$\pm$0.02 & 14.54$\pm$0.01 &  20.5$\pm$0.9 &  17.2$\pm$0.8 \\
0500+019  &  33 & 26 & 42 & 26 & 19.30$\pm$0.03 & 16.94$\pm$0.02 & 20.43$\pm$0.10& 18.09$\pm$0.10 \\
0710+439  &  16 & 16 & 27 & 15 & 18.26$\pm$0.02 & 16.29$\pm$0.01 & 20.06$\pm$0.16& 18.17$\pm$0.17 \\
3C 200    &   8 &  6 & 13 &  5 & 17.70$\pm$0.01 & 15.85$\pm$0.01 &  20.6$\pm$0.4 &  19.0$\pm$0.4 \\
0941$-$080  &   5 &  4 &  9 &  2 & 16.92$\pm$0.01 & 15.26$\pm$0.01 &  20.3$\pm$0.5 &  19.6$\pm$1.0 \\
0941$-$080b\tablenotemark{b}&  9 &  1 & 22 &  1 & 17.62$\pm$0.02 & 16.34$\pm$0.02 &    23$\pm$2   &    21$\pm$2 \\
1323+321  &  16 & 14 & 16 &  7 & 18.07$\pm$0.02 & 16.07$\pm$0.01 & 20.07$\pm$0.18& 18.88$\pm$0.33 \\
3C 268.3  &  42 & 17 & 66 & 44 & 18.18$\pm$0.02 & 16.11$\pm$0.01 & 19.90$\pm$0.15& 16.40$\pm$0.06 \\
3C 303.1  &   5 &  1 & 10 &  3 & 16.89$\pm$0.01 & 15.38$\pm$0.01 &    23$\pm$2   &  19.1$\pm$0.6 \\
3C 456    &  10 &  6 & 30 & 14 & 17.08$\pm$0.01 & 15.36$\pm$0.01 &  20.0$\pm$0.4 & 17.37$\pm$0.18 \\
3C 460    &   4 &  1 & 10 &  1 & 17.19$\pm$0.01 & 15.33$\pm$0.01 &    23$\pm$2   &    21$\pm$2 \\
3C 460b\tablenotemark{b}&  7 &  1 & 17 &  6 & 18.25$\pm$0.02 & 17.11$\pm$0.03 &    23$\pm$2   &  20.9$\pm$0.4 \\
2342+821  &  34 & 37 & 54 & 26 & 18.11$\pm$0.02 & 16.58$\pm$0.01 & 18.70$\pm$0.07& 17.69$\pm$0.10 \\
2352+495  &   7 &  4 & 15 &  1 & 16.70$\pm$0.01 & 15.15$\pm$0.01 &  20.2$\pm$0.5 &    21$\pm$2 \\
\enddata
\tablenotetext{a}{Maximum luminosity fraction attributable to unresolved nuclear component
(\%$^{\mbox{max}}_{\mbox{psf}}$), based on fits to the peak flux; and modeled fraction (\%$^{\mbox{act}}_{\mbox{psf}}$) 
as described in the text. Fractional uncertainties are large, and should be taken on the
order of a few percent.}
\tablenotetext{b}{Interacting companion galaxy.}
\end{deluxetable}

\begin{deluxetable}{llllclllc}
\tablenum{4}
\tablecaption{Profile fitting results}
\tablehead{
  \colhead{} & \multicolumn{4}{c}{F205W\tablenotemark{a}} & 
  \multicolumn{4}{c}{F110W\tablenotemark{a}}
\\
  \cline{2-5}
  \cline{6-9}
\\
  \colhead{Source}  & 
  \colhead{R$^{\rm n}_e$} & \colhead{$\mu^{\rm n}_e$} & 
  \colhead{$\mu_0$} & \colhead{n} & 
  \colhead{R$^{\rm n}_e$} & \colhead{$\mu^{\rm n}_e$} & 
  \colhead{$\mu_0$} & \colhead{n} 
}
\startdata
3C 16    &3.31$\pm$0.25 & 18.37$\pm$0.17 & 15.97 &  3.70$\pm$0.57 & 2.22$\pm$0.07 & 19.47$\pm$0.07 & 17.57 &  2.26$\pm$0.13 \\
3C 19    &4.17$\pm$0.37 & 18.46$\pm$0.21 & 16.17 &  2.96$\pm$0.51 & 3.23$\pm$0.07 & 19.90$\pm$0.05 & 17.72 &  2.05$\pm$0.10 \\
3C 42    &8.82$\pm$0.50 & 19.51$\pm$0.12 & 15.64 &  4.06$\pm$0.57 & 5.08$\pm$0.14 & 20.41$\pm$0.06 & 17.21 &  3.24$\pm$0.17 \\
3C 46    &16.4$\pm$0.91 & 20.57$\pm$0.11 & 15.85 &  6.19$\pm$0.80 & 3.19$\pm$0.09 & 19.46$\pm$0.07 & 17.57 &  1.52$\pm$0.08 \\
3C 46b\tablenotemark{b}&2.56$\pm$0.05 & 17.49$\pm$0.05 & 15.76 &  1.66$\pm$0.08 & 3.32$\pm$0.12 & 19.75$\pm$0.09 & 17.23 &  3.05$\pm$0.21 \\
3C 67    &1.68$\pm$0.10 & 16.48$\pm$0.17 & 14.36 &  3.09$\pm$0.40 & 1.36$\pm$0.05 & 17.75$\pm$0.11 & 15.65 &  2.19$\pm$0.14 \\
3C 93.1  &4.41$\pm$0.13 & 17.85$\pm$0.07 & 14.14 &  3.67$\pm$0.29 & 4.96$\pm$0.30 & 20.01$\pm$0.13 & 15.79 &  6.18$\pm$0.77 \\
0404+768 &3.18$\pm$0.42 & 17.38$\pm$0.31 & 14.52 &  7.54$\pm$2.67 & 2.62$\pm$0.17 & 19.31$\pm$0.17 & 16.84 &  2.33$\pm$0.27 \\
0428+205 &8.81$\pm$0.30 & 19.14$\pm$0.07 & 14.68 &  4.39$\pm$0.40 & 4.46$\pm$0.20 & 19.92$\pm$0.10 & 16.40 &  2.80$\pm$0.30 \\
0500+019 &3.11$\pm$0.45 & 17.73$\pm$0.30 & 14.94 &  5.24$\pm$1.31 & 2.97$\pm$0.22 & 20.10$\pm$0.17 & 17.40 &  3.04$\pm$0.38 \\
0710+439 &4.52$\pm$0.27 & 18.11$\pm$0.14 & 14.94 &  3.57$\pm$0.46 & 3.04$\pm$0.11 & 19.40$\pm$0.09 & 16.69 &  2.33$\pm$0.15 \\
3C 200   &4.18$\pm$0.22 & 17.71$\pm$0.13 & 15.15 &  3.86$\pm$0.48 & 3.39$\pm$0.04 & 19.01$\pm$0.03 & 16.82 &  1.34$\pm$0.05 \\
0941$-$080 &4.22$\pm$0.11 & 17.98$\pm$0.06 & 15.02 &  2.45$\pm$0.15 & 3.07$\pm$0.08 & 18.98$\pm$0.07 & 16.20 &  1.79$\pm$0.11 \\
0941$-$080b\tablenotemark{b}&1.48$\pm$0.06 & 16.79$\pm$0.12 & 15.14 &  2.47$\pm$0.23 & 1.70$\pm$0.14 & 18.69$\pm$0.20 & 16.39 &  9.15$\pm$1.27 \\
3C 268.3 &0.41$\pm$0.21 & 12.75$\pm$1.37 & 13.41 &  10$\pm$7      & 0.46$\pm$0.21 & 15.65$\pm$1.20 & 15.88 &  10$\pm$7      \\
1323+321 &4.30$\pm$0.13 & 18.14$\pm$0.08 & 15.25 &  2.41$\pm$0.18 & 2.45$\pm$0.05 & 18.99$\pm$0.05 & 16.47 &  2.71$\pm$0.09 \\
3C 303.1 &4.63$\pm$0.12 & 18.11$\pm$0.06 & 14.96 &  3.00$\pm$0.19 & 3.58$\pm$0.06 & 19.15$\pm$0.04 & 16.42 &  2.21$\pm$0.08 \\
3C 456   &1.43$\pm$0.09 & 15.78$\pm$0.16 & 13.82 &  5.78$\pm$0.81 & 1.98$\pm$0.03 & 18.25$\pm$0.04 & 15.82 &  2.47$\pm$0.08 \\
3C 460   &4.18$\pm$0.12 & 17.92$\pm$0.07 & 15.00 &  3.36$\pm$0.24 & 3.12$\pm$0.05 & 19.14$\pm$0.04 & 16.72 &  2.22$\pm$0.07 \\
3C 460b\tablenotemark{b}&2.04$\pm$0.07 & 17.78$\pm$0.10 & 16.10 &  1.35$\pm$0.09 & 2.73$\pm$0.28 & 19.84$\pm$0.23 & 17.38 &  2.45$\pm$0.48 \\
2342+821 &1.41$\pm$0.18 & 15.53$\pm$0.40 & 14.39 &  3.23$\pm$0.83 & 0.44$\pm$0.05 & 14.49$\pm$0.34 & 15.72 &  6.90$\pm$1.32 \\
2352+495 &2.52$\pm$0.06 & 16.93$\pm$0.06 & 14.42 &  4.01$\pm$0.23 & 2.19$\pm$0.03 & 18.32$\pm$0.03 & 15.86 &  2.60$\pm$0.07 \\
\enddata
\tablenotetext{a}{R$^{\rm n}_e$ is the effective radius (in kpc) of the applied 
r$^\frac{1}{n}$-law, $\mu^{\rm n}_e$ is the corresponding effective surface
brightness, and $\mu_0$ is the observed central surface brightness within a
1 pixel radius of the nucleus. The exponent of the best fitting law is given
by n.}
\tablenotetext{b}{Interacting companion galaxy.}
\end{deluxetable}

\clearpage

\begin{figure*}[t]
\epsscale{2.2}
\vspace{-2cm}
\caption{3C 16 (\ion{FR}{2}), $z=0.405$. NICMOS $J$ and $K$ band images
(F110W and F205W, top two panels). The lower left panel presents the
$J-K$ color map, with the greyscale running from red ($-0.2$) to
neutral to blue ($+0.2$). The F205W wide field image is shown in the
lower right panel, with the radio symmetry axis and size indicated by
the lines. If the radio structure is larger than the panel, the arrows
point outwards; if the radio size is too small to be plotted
meaningfully, the arrows are pointed inwards. Galaxy model subtraction
residuals are shown in the small boxes at the bottom.}
\label{figure10}
\end{figure*}

\begin{figure*}[t]
\caption{3C 19 (\ion{FR}{2}, $z=0.482$)}
\label{figure11}
\end{figure*}

\begin{figure*}[t]
\caption{3C 42 (\ion{FR}{2}, $z=0.395$)}
\label{figure12}
\end{figure*}

\begin{figure*}[t]
\caption{3C 46 (\ion{FR}{2}, $z=0.437$)}
\label{figure13}
\end{figure*}

\begin{figure*}[t]
\caption{3C 67 (CSS, $z=0.310$)}
\label{figure14}
\end{figure*}

\begin{figure*}[t]
\caption{3C 93.1 (CSS, $z=0.243$)}
\label{figure15}
\end{figure*}

\begin{figure*}[t]
\caption{0404+768 (GPS/CSO, $z=0.599$)}
\label{figure16}
\end{figure*}
\clearpage

\begin{figure*}[t]
\caption{0428+205 (GPS/CSO, $z=0.219$)}
\label{figure17}
\end{figure*}

\begin{figure*}[t]
\caption{0500+019 (GPS/CSO, $z=0.583$)}
\label{figure18}
\end{figure*}

\begin{figure*}[t]
\caption{0710+439 (GPS/CSO, $z=0.518$)}
\label{figure19}
\end{figure*}

\begin{figure*}[t]
\caption{3C 200 (\ion{FR}{2}, $z=0.458$)}
\label{figure20}
\end{figure*}

\begin{figure*}[t]
\caption{0941$-$080 (GPS, $z=0.228$)}
\label{figure21}
\end{figure*}

\begin{figure*}[t]
\caption{3C 268.3 (CSS, $z=0.371$)}
\label{figure22}
\end{figure*}

\begin{figure*}[t]
\caption{1323+321 (GPS, $z=0.369$)}
\label{figure23}
\end{figure*}

\begin{figure*}[t]
\caption{3C 303.1 (CSS, $z=0.267$)}
\label{figure24}
\end{figure*}

\begin{figure*}[t]
\caption{3C 456 (\ion{FR}{2}, $z=0.233$)}
\label{figure25}
\end{figure*}

\begin{figure*}[t]
\caption{3C 460 (\ion{FR}{2}, $z=0.268$)}
\label{figure26}
\end{figure*}

\begin{figure*}[t]
\caption{2342+821 (GPS/CSO, $z=0.735$)}
\label{figure27}
\end{figure*}

\begin{figure*}[t]
\caption{2352+495 (GPS/CSO, $z=0.237$)}
\label{figure28}
\end{figure*}

\end{document}